\documentclass[a4paper, 10pt]{article}
\usepackage[margin=1.2in]{geometry}

\usepackage{titling}

\usepackage{authblk}
\usepackage{amsmath,amssymb}
\usepackage{xfrac}
\usepackage{float}
\usepackage{xcolor}
\usepackage{nicefrac}
\usepackage{graphicx}

\usepackage{parskip}
\title{Dendrites with corners}
\author[1]{E.S. Nani\thanks{Corresponding author e-mail address: sumanth-nani.enugala@polytechnique.edu}} \author[1]{T. Philippe} \author[1]{M. Plapp} \affil[1]{\normalsize Laboratoire de Physique de la Mati\`ere Condens\'ee, CNRS, Ecole Polytechnique, Institut Polytechnique de Paris, 91120 Palaiseau, France}

\date{30/11/2023}

\begin{document}
\vspace{-4cm}
\maketitle 

\begin{abstract}
A phase-field model for diffusion-limited crystal growth is formulated that is capable of handling highly anisotropic interfaces. It uses a Willmore regularization that yields corners of finite size. An asymptotic analysis reveals that Herring's law is recovered for the advancing surfaces. The model is validated by conducting simulations of dendritic growth for low anistorpies and comparing the results to the data from the literature. The model makes it possible to simulate high anisotropy dendrites for which the standard phase-field models are ill-posed. In this regime, the interplay between a Herring instability on the dendrite flanks and the corner regularization creates zig-zag shaped corrugations and leads to a non-monotonic trend of tip velocity as a function of anisotropy strength. 
\end{abstract}

\section{Introduction}

Crystal growth from a melt can produce a large variety of complex morphologies. In solidification, the resulting microstructures determine many material properties. The phase-field method has been hugely successful in capturing this pattern formation process in numerical simulations.

Dendrites are one of the most widespread solidification microstructures. Their morphology is determined by the anisotropy of the solid-liquid interfacial free energy. Phase-field studies of dendritic growth have mostly been limited to low and moderate anisotropy strengths. This is the relevant case for metals which have atomically rough interfaces. However, for substances which have atomically smooth (facetted) surfaces, the interfacial energy anisotropy is much stronger~\cite{jackson1958liquid,fisher1998fundamentals}.

When the interfacial anisotropy becomes so strong that certain surface orientations are excluded from the equilibrium shape, the equations of motion need to be regularized both in the sharp-interface and phase-field treatments~\cite{herring1951some,torabi2009new}. A first regularization technique for phase-field models was proposed by Eggleston et al.~\cite{eggleston2001phase}, but needs to be adapted to each functional form of the interfacial anisotropy. More recently, a Willmore regularization was introduced in phase-field models for surface diffusion~\cite{torabi2009new,salvalaglio2015faceting} and kinetically-limited crystal growth ~\cite{loreti2000propagation,philippe2020regularized}. This regularization creates corners of fixed size $\alpha$ which bridge the forbidden orientations. Here, we use the combination of the latter model with the grand-potential formulation of alloy solidification\cite{plapp2011unified,choudhury2012grand} to study diffusion-limited dendritic growth with anisotropy of arbitrary strengths. An asymptotic analysis enables us to make rigorous connection to the relevant sharp-interface problem.

In the next section, the regularized model is presented and benchmarked. Simulation results generated using the model are reported and discussed in section after before concluding the article.

\section{Model Construction and Benchmarks}

\subsection{Model}

The standard formulation of phase-field models is limited to moderate anisotropies with positive interface stiffness for all orientations. To extend it to larger anisotropies, we add the Willmore regularization proposed in Refs.~\cite{torabi2009new,philippe2020regularized}. The core idea of this regularization is to mimic the approach taken in the sharp-interface theory, namely, to penalize curvatures in the evolving crystal shape so as to avoid sharp corners. To this end, a new term is added to the functional of the grand-potential model for isothermal alloy solidification~\cite{plapp2011unified}, which becomes
\begin{eqnarray}
\Omega & = & \int \left[ ( \omega_{\text{sol}} (\mu) - \omega_{\text{liq}} (\mu) ) g(\phi) + \omega_{\text{liq}}(\mu) + \frac{ \gamma(\theta(\nabla \phi)) } {\epsilon} \left( f_{\text{dw}} (\phi) + \frac{\epsilon^2}{2} (\nabla \phi)^2 \right) \right.\nonumber \\
   & & \left.\mbox{} +  \gamma_0 \frac{\alpha^2}{2 \epsilon} \left(\frac{f'_{\text{dw}} (\phi)}{\epsilon} - \epsilon \Delta \phi \right)^2  \right] dV
\end{eqnarray}
where $\omega_{\text{sol}}$ and $\omega_{\text{liq}}$ are the volumetric grand-potential densities of the crystal and the liquid phases, respectively; $\mu$ is the diffusion potential; $\phi$ is the phase-indicator marking liquid and crystal phases with the its values of $0$ and $1$, respectively; $g(\phi)$ is a function interpolating between $0$ and $1$ as $\phi$ ranges over $[0,1]$; $f_\text{dw}(\phi)$ is the double-well potential; $\gamma(\theta(\nabla \phi))$ is the interfacial energy as a function of orientation $\theta$ which in turn is specified using $\nabla \phi$; $\gamma_0$ is the average of $\gamma(\theta)$; and $\epsilon$ is the interface-width specifier. The bracketed expression in the last term of the functional is the diffuse interface approximation of the local interface curvature. Hence this term penalizes curvatures of either signs and creates smoothed corners of fixed size controlled by the Willmore regularization parameter $\alpha$. In contrast to just adding a $\left( \Delta \phi \right)^2$ term~\cite{wise2007solving,wheeler2006phase}, this regularization preserves the hyperbolic tangent interface profile known from the standard phase-field models.

The governing equations derived from this functional are as follows:
\begin{subequations} \label{eq:Model_Equations}
\begin{align} \label{eq:Main_Model_phiEquation}
\tau \frac{\partial \phi}{\partial t} = &  -\frac{1}{\epsilon}  \left(\omega_{\text{sol}} (\mu) -\omega_{\text{liq}}(\mu) \right)  g'(\phi)      -  \frac{1}{\epsilon^2}  \gamma(\theta(\nabla \phi))  f_{\text{dw}}'(\phi) \nonumber \\
& + \nabla \cdot \left[\gamma(\theta(\nabla \phi)) \nabla \phi \right] + \frac{1}{\epsilon^2} \nabla \cdot \left[ \left(f_{\text{dw}}(\phi) + \frac{\epsilon^2}{2} |\nabla \phi|^2 \right)  \nabla_{\nabla \phi} \left( \gamma(\theta(\nabla \phi)) \right) \right] \nonumber \\
& + \gamma_0 \frac{\alpha^2}{\epsilon^4}  \left( \epsilon^2 \nabla^2 \phi - f_{\text{dw}}'(\phi)  \right)  f_{\text{dw}}''(\phi)  - \gamma_0 \frac{\alpha^2}{\epsilon^2}  \nabla^2 \left(\epsilon^2 \nabla^2\phi - f_{\text{dw}}'(\phi) \right) 
\end{align}
\begin{align} \label{eq:Main_Model_muEquation}
\frac{\partial \mu}{\partial t} = \left[ \nabla \cdot \left( \left[ \left(D_{\text{sol}} - D_{\text{liq}}\right) h(\phi) + D_{\text{liq}} \right] \nabla \mu \right) - \frac { \left( c_{\text{sol}} (\mu) - c_{\text{liq}} (\mu) \right) }{c,_{\mu}} f'(\phi) \frac{\partial \phi}{\partial t} \right]
\end{align}
\end{subequations}
where $D_\text{sol}$ and $D_\text{liq}$ are the diffusion coefficients, and $c_\text{sol}$ and $c_\text{liq}$ are the compositions of the solid and the liquid phases, respectively. The symbol $c,_{\mu}$ is used to denote $ \left( \frac{d c_{\text{sol}}}{d \mu} - \frac{d c_{\text{liq}}}{d \mu} \right) g(\phi) + \frac{d c_{\text{liq}}}{d \mu} $. $h(\phi)$ and $f(\phi)$ are interpolation functions like $g(\phi)$, and $\tau$ is an inverse mobility parameter.

Matched asymptotic analysis demonstrates that the generalized Stefan problem is recovered by the model of Eq.~\eqref{eq:Model_Equations} \iffalse(in the limit of vanishing $\epsilon$)\fi except for the Gibss-Thomson relation, which is replaced by Herring's law:
\begin{align}\label{eq:Chemical_Herrings_Law}
 \tilde{\tau} v_n = \left(\omega_{\text{liq}} (\mu) -\omega_{\text{sol}}(\mu) \right) - \left( \gamma(\theta) + \gamma''(\theta) \right)\kappa + \gamma_0 \alpha^2 \left( \frac{\kappa^3}{2}  + \kappa_{ss} \right) 
\end{align}
where $v_n$ and $\kappa$ are respectively the normal velocity and the curvature of the solid-liquid boundary; $\kappa_{ss}$ is the surface Laplacian of curvature. $\tilde{\tau}$ is a constant that describes the strength of interface attachment kinetics.

In the current article, the focus is on diffusion-limited growth. Hence, we choose the phase-field relaxation parameter $\tau$ such that the interface attachment kinetics $\tilde{\tau}$ is zero, following methods analogous to those developed by Almgren~\cite{almgren1999second}. In this preliminary study, we use equal diffusivities in the liquid and the solid ($D_\text{sol} = D_\text{liq})$ and denote the common value by $D$.

\subsection{Dendritic growth for low anisotropies}

We tested Eq.~\eqref{eq:Chemical_Herrings_Law} by standard benchmarks such as kinetically-limited and diffusion-limited planar front growth and critical nucleus evolution, with excellent success. For low anisotropies where the regularization terms are not required ($\alpha = 0$), as a first non-trivial benchmark problem, we simulated dendritic growth for low anisotropies. An anisotropic crystal seed growing at the expense of a supersaturated infinite matrix develops into a dendritic pattern and eventually enters a steady-state.
We compared the simulated operating state with the results of Karma-Rappel~\cite{karma1998quantitative}. 

We chose the following forms for the various functions: $f_{\text{dw}}(\phi) = 18 \phi^2 (1-\phi)^2$, $g(\phi)=\phi^3(10-15\phi+6\phi^2)$, $f(\phi)=\phi$, $\omega_{\text{sol}}(\mu) = -\mu^2/4A - \mu c_{\text{sol}}^{\text{eq}}$, $\omega_{\text{liq}}(\mu) = -\mu^2/4A - \mu c_{\text{liq}}^{\text{eq}}$, and 
\begin{equation}
\gamma(\theta)=\gamma_0 \left(1+\epsilon_4 \cos(4 \theta) \right). 
\end{equation}
These choices relate the various thermo-physical parameters of the current model to those of Karma-Rappel's in the following manner: supersaturation $\Delta$ matches up with $\frac{\left(c_{\text{liq}}^{\text{eq}} - c_0 \right)}{\left(c_{\text{liq}}^{\text{eq}}  - c_{\text{sol}}^{\text{eq}} \right)}$ where $c_0$ is the liquid's far-field as well as the initial composition, the interface width $2\sqrt{2}W_0$ scales as $2\epsilon/3$, the capillary length $d_0$ coincides with $\frac{\gamma_0}{2A\left(c_{\text{sol}}^{\text{eq}} - c_{\text{liq}}^{\text{eq}}\right)^2}$, $u$ of Ref.~\cite{karma1998quantitative} corresponds to $\frac{\mu}{2A \left(c_{\text{sol}}^{\text{eq}} - c_{\text{liq}}^{\text{eq}}\right)}$, and finally, the diffusion coefficient $D$ and the anisotropy strength $\epsilon_4$ are the same across both the treatments. 

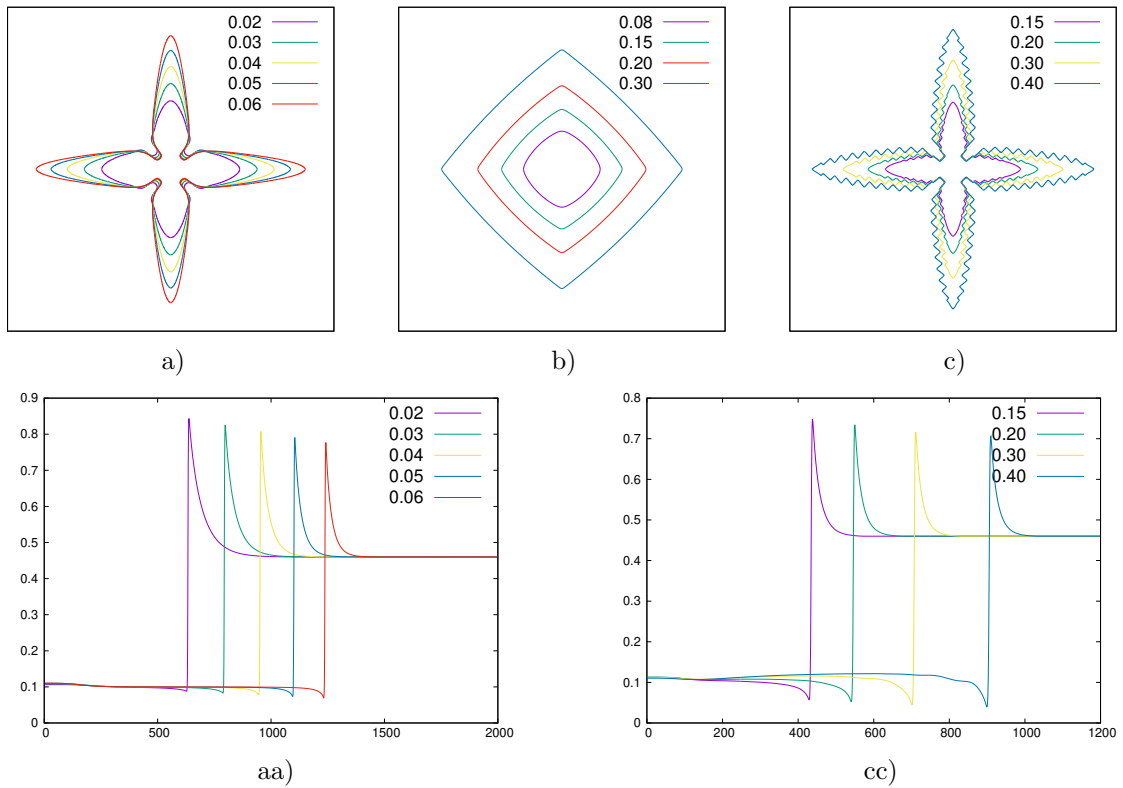
\begin{figure}[!h]
\begin{center}
\includegraphics[width=1.0\textwidth]{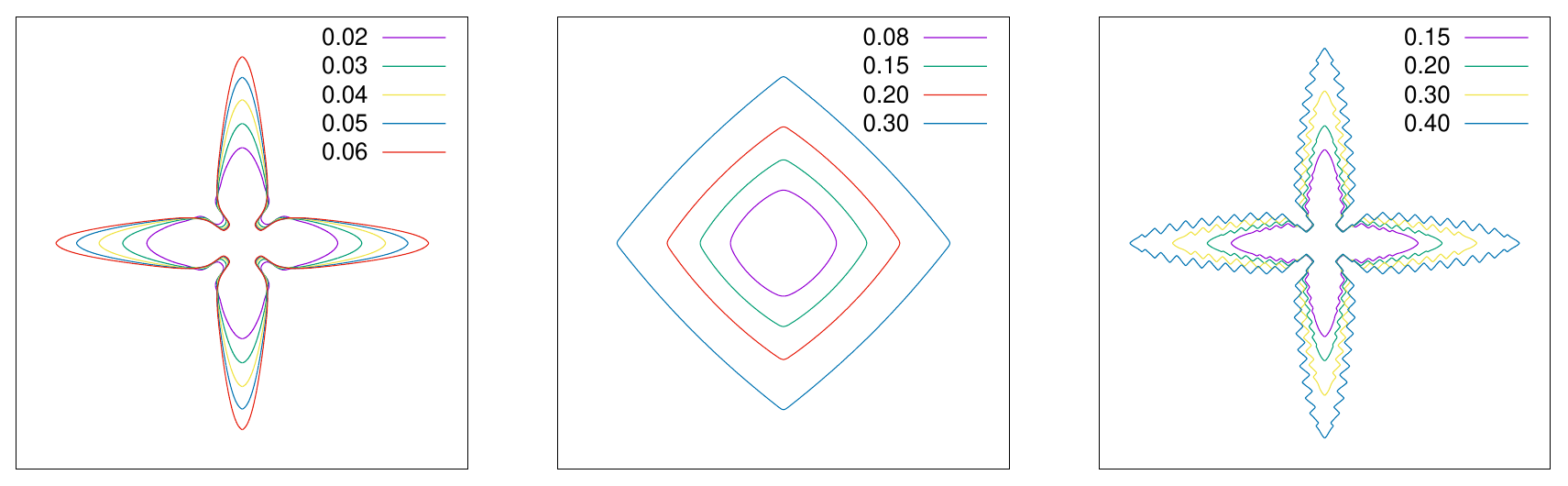}
\put(-362,-10){a)}
\put(-216,-10){b)}
\put(-69,-10){c)}\\
\includegraphics[width=1.0\textwidth]{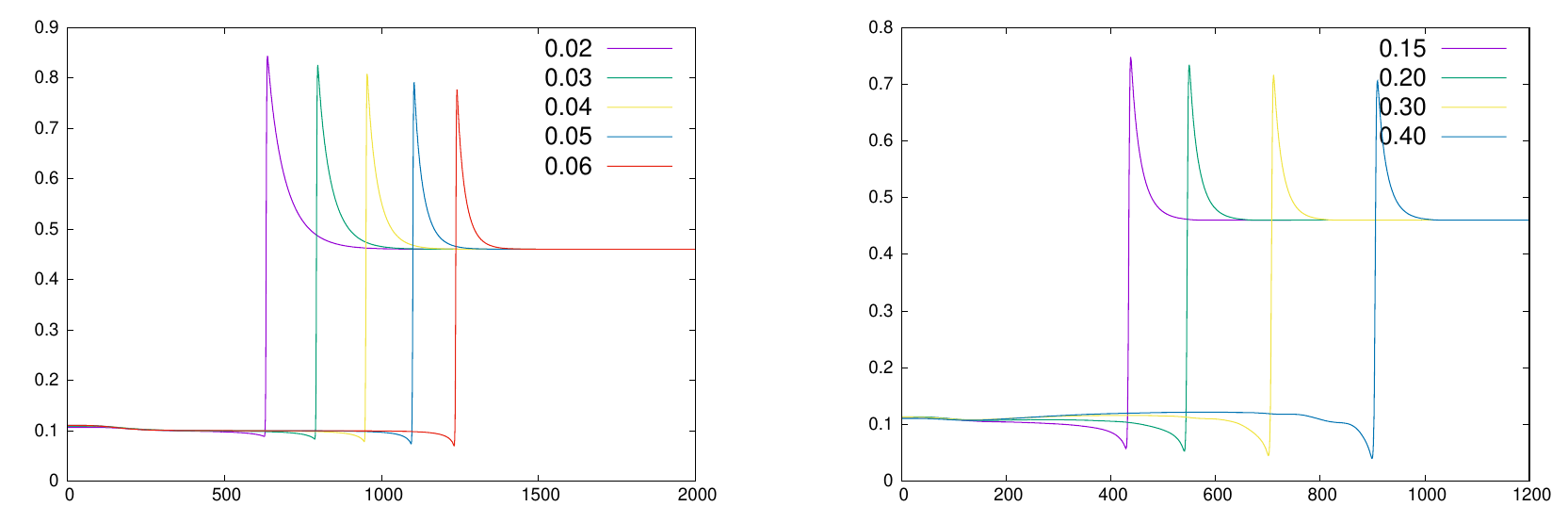}
\put(-326,-10){aa)}
\put(-99,-10){cc)}
\caption{Results obtained using the presented model (Eq.~\eqref{eq:Model_Equations}). The plots show the $\phi=0.5$ contours of the simulations for various anisotropy strengths for a) the traditional dendritic growth, i.e., with $\alpha=0$, b) growth under constant supersaturation, i.e., by holding $\mu$ constant, and c) non-zero regularization parameter. Also shown are the concentration profiles along the central axis of the rightward growing dendrite arms of the cases a) and b) in sub-figures aa) and cc), respectively.} \label{fig:Benchmarks}
\end{center}
\end{figure}

By keeping the set $\{c_{\text{liq}}^{\text{eq}}, c_{\text{sol}}^{\text{eq}}, A \}$ fixed, physical situations corresponding to various $\Delta-W_0-d_0$ combinations of \textrm{TABLE II} of Karma-Rappel are simulated by choosing the $c_0-\epsilon-\gamma_0$ triplets accordingly. The spatial discretizations and the grid sizes chosen for each simulation are exactly identical to those listed in the table. The recovered steady-state dendrite tip speeds are within $2-7\%$ of the expected values for all the cases tested. Growth profiles simulated for various anisotropy strengths $\epsilon_4$ using the parameter set of Table~\ref{tab:infile_parameters}, which corresponds to the dimensionless parameter set of $\Delta=0.55$, $D=4.0$ and $d_0 = 0.1385$ of Karma and Rappel, are reported in Fig.~\ref{fig:Benchmarks}a). The starting seed size is the same for all the trials. The depicted interface profiles also correspond to the same but a later timestep. That means, the trend of dendrite tips becoming faster and sharper as anisotropy strength is increased can be witnessed to be correctly reproduced by our simulations. Curves in Fig.~\ref{fig:Benchmarks}aa) correspond to concentration profiles along the dendrite axis. The concentration in the liquid next to the tip decreases with increasing anisotropy, which is of course consistent with the Gibbs-Thomson effect.

\begin{table}[!htbp]
\centering
\caption{Table showing the material, modeling and simulation parameter set used for benchmarking the model.}
\begin{tabular}{c c c c c c c c c }
\hline 
$c_{\text{liq}}^{\text{eq}}$ & $c_{\text{sol}}^{\text{eq}}$ & $A$  & $\gamma_0$ & $D$ &  Grid discretization & $\epsilon$ & $\tau$ & $\Delta t$   \\
0.9  & 0.1  & 0.78125 &  0.1385   &  4.0 & $\Delta x = \Delta y = 0.2$ & $7.5 \sqrt{2} \Delta x$ & $\epsilon \frac{\left(c_{\text{sol}}^{\text{eq}} - c_{\text{liq}}^{\text{eq}}\right)^2}{\frac{D}{2A}} \frac{47}{360}$ & $0.2 \frac{(\Delta x)^2}{D}$  \\
\hline
\end{tabular}
\label{tab:infile_parameters}
\end{table}

\subsection{Growth under constant driving force}

In order to benchmark the regularization terms, we switched off the coupling to transport, started with an initial circular seed of $5$ grid cells, and let it grow under the drive of a constant supersaturation until it occupies a certain percentage of the simulation domain, at which point, volume preservation is imposed for further evolution by a Lagrange multiplier implementation. The material and simulation parameter-sets used are identical to those above. Results for various anisotropy strengths and $\alpha$ held fixed at $0.5$ (in the same units) are reported in Fig.~\ref{fig:Benchmarks}b). Except for corners, these contours were seen to be in excellent agreement with the respective Wulff shapes.

\section{Results and Discussion}

\begin{figure}[!h]
\begin{center}
\includegraphics[width=\textwidth]{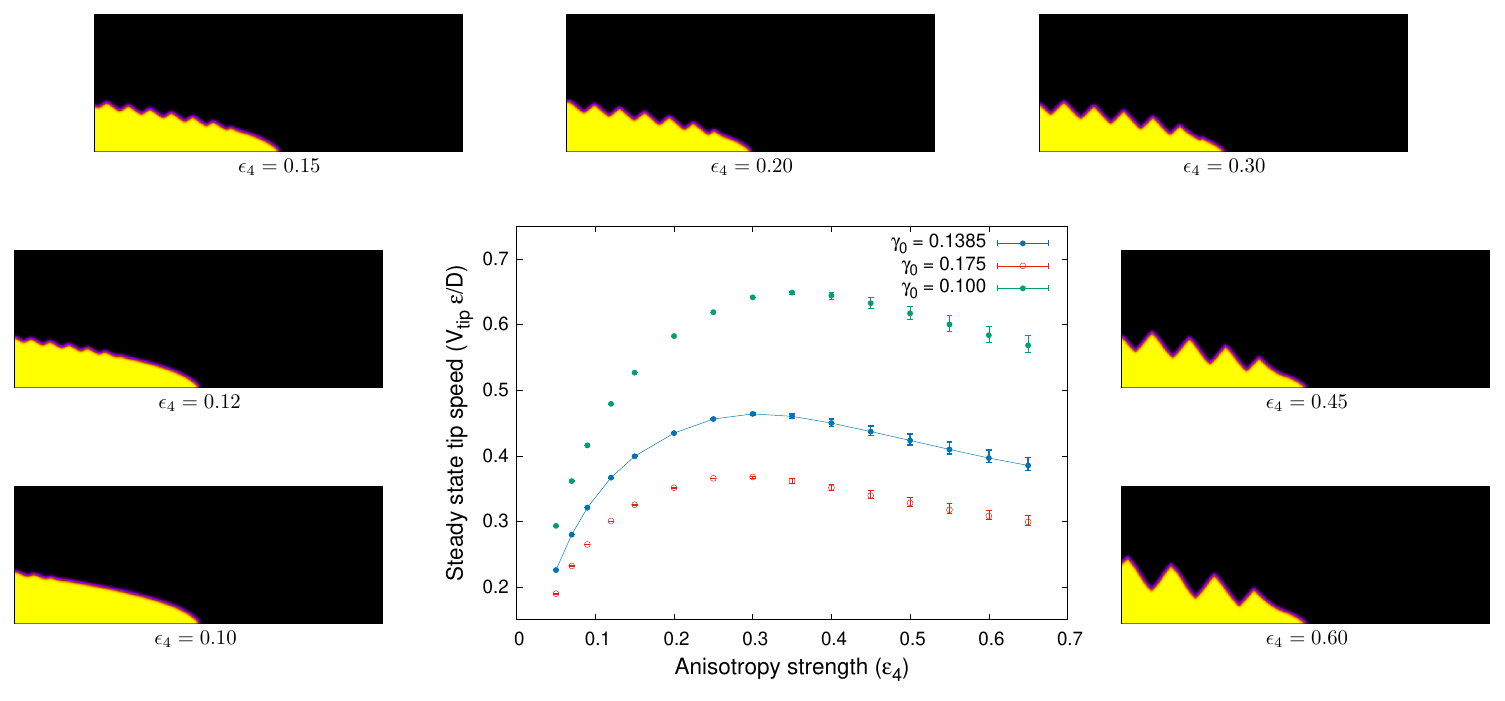}
\caption{Crystal growth morphologies and steady-state tip velocities predicted by the present model for various anisotropy strengths $\epsilon_4$.}  \label{fig:Main_Results}
\end{center}
\end{figure}

For $\epsilon_4 > \sfrac{1}{15}$, ranges of interface orientations around $\theta=0$, $\theta = \pm \pi /2$, and $\theta=\pi$ become unstable, and the standard phase-field model becomes ill-posed. With our regularized model, however, dendrites can be simulated for $\epsilon_4$ larger than $\sfrac{1}{15}$. This is illustrated in Figs.~\ref{fig:Benchmarks}c) and~\ref{fig:Main_Results} for a value of $\alpha=3.61 \times d_0$ and otherwise the same parameter set used for Fig~\ref{fig:Benchmarks}a).

Fig.~\ref{fig:Benchmarks}c) shows growth contours of crystals for various anisotropies evolved for different times to facilitate better visualization. The typical dendritic shape persists even for large anisotropies, which indicates that growth is still diffusion-limited. Corrugations of zig-zag shape appear behind the tip for large enough anisotropy strengths. Fig.~\ref{fig:Benchmarks}cc) shows that the concentration fields remain smooth even for high anisotropies.

For a quantitative evaluation of the tip operating state, one arm is followed for long times in a moving simulation box. The tip morphologies are illustrated in the snapshot pictures in Fig~\ref{fig:Main_Results}, and the steady-state tip velocity is given by the solid curve (with error bars) in the graph. The tip growth speed first increases as a function of anisotropy strength, then passes through a maximum, and slowly decreases beyond. Moreover, the tip velocity, which is constant for low anisotropies, starts to oscillate beyond the maximum. The error bars in Fig.~\ref{fig:Main_Results} indicate the minimum and maximum velocities of the oscillation cycles. These observations can be interpreted as follows.

Three different regimes can be identified. For $\epsilon_4$ smaller than $\sfrac{1}{15}$, the dendrite is smooth everywhere, and the regularization only slightly modifies the anisotropic needle crystal shape. Therefore, it is not surprising to recover the familiar behavior of increasing of growth speed with anisotropy strength. This trend continues for a range of anisotropies beyond $\epsilon_4 = \sfrac{1}{15}$. In this regime, the dendrite tip is a corner, but the dendrite flanks remain smooth for a considerable distance behind the tip. Eventually, corrugations start to develop on the dendrite flank, because the interface enters an unstable orientation range around $\theta = \pi/2$. Since with increasing anisotropy strength the unstable ranges get larger, the corrugations develop closer and closer to the tip until these undulations induce an oscillation of the tip speed itself. The velocity oscillations can likely be explained by the modification of the diffusion field created by these zig-zags.

While the corrugations may seem reminiscent of the sidebranches in solidfication dendrites, they are created by an altogether different mechanism. Whereas sidebranches result from a selective amplification of thermal noise along the flanks of a dendrite~\cite{langer1987dendritic,karma1998quantitative}, here the corrugations are created by a Herring instability that only sets in when the interface orientation reaches an unstable range. In our simulations, this instability is triggered by the time-dependent dendrite shape without the explicit addition of the thermal fluctuations.

For a comprehensive understanding, the dependence of the operating state on all of the parameters should be investigated. As a first step in this direction, simulations were repeated for two more values of the surface energy $\gamma_0$, viz. $\gamma_0=0.175$ and $\gamma_0=0.100$ corresponding to the red and green data points of Fig.~\ref{fig:Main_Results}, respectively. As expected from the standard scaling theory of dendritic growth, lowering the surface energy makes dendrites sharper and faster. However, the tip velocity does not exactly follow the scaling $v \propto \nicefrac{1}{\gamma_0}$ predicted by solvability theory~\cite{barbieri1987velocity}. This is due to the presence of the corner size $\alpha$ as an additional length scale in the problem.

The present simulations have remained limited to the symmetric model $\left( D_{\text{sol}} = D_{\text{liq}}  \right)$. However, solvability theory for low anisotropy dendrites predicts that unequal diffusivities do only modify the selection constant, whereas the qualitative behavior remains the same~\cite{barbieri1989predictions}. There is no reason to assume that it would be different for high anisotropy dendrites. Hence we expect that our main results, namely the zig-zagged corrugations and the non-monotonic trend in the tip velocities, should not be affected by the diffusivity ratio.

\section{Conclusion}

We have demonstrated that our regularized phase-field model can be used to explore the physics of dendritic growth at large anisotropies. The presence of a stable corner on the dendrite tip leads to a continuation of needle crystal growth beyond the limit $\epsilon_4 = \sfrac{1}{15}$. For a better theoretical understanding of these new solutions, we are presently exploring an extension of the linearized microscopic solvability theory~\cite{barbieri1989predictions} to the current problem, namely the generalized Stefan problem with the Gibbs-Thomson condition replaced by the Herring's equation.

\section*{Acknowledgement}
The authors gratefully acknowledge the financial support provided by Agence Nationale de la Recherche, France, under the project FACET.

%\bibliographystyle{unsrt}
%\bibliography{references}

\end{document}